\documentclass[a4paper]{article}

\usepackage{icrc2013}
\usepackage{graphicx}
\usepackage{graphics}
\usepackage{placeins}

\title{Calibration of the ASTRI SST-2M Prototype using Muon Ring Images.}

\shorttitle{ASTRI SST-2M Prototype: Calibration with Muons}

\authors{
Elisabetta Strazzeri$^{1}$,
Giacomo Bonnoli$^{2}$,
Saverio Lombardi$^{3}$,
Maria Concetta Maccarone$^{1}$,\\
Teresa Mineo$^{1}$, for the ASTRI Collaboration$^{4}$.
}

\afiliations{
$^1$ INAF, IASF-Palermo, Via Ugo La Malfa 153, I-90146 Palermo, Italy\\
$^2$ INAF, Osservatorio Astronomico di Brera, via E. Bianchi 46, 23807 Merate, Italy\\
$^3$ INAF, Osservatorio Astronomico di Roma, via di Frascati 33 - 00044 Monteporzio Catone, Italy\\
$^4$ http://www.brera.inaf.it/astri/
}

\email{Elisabetta.Strazzeri@ifc.inaf.it}

\abstract{The study of ring images generated from high-energy muons is a very useful tool for the performance monitoring and calibration of any Imaging Atmosphere Cherenkov Telescope. Isolated muons travelling towards the telescope light collector system produce characteristic Cherenkov ring images in the focal plane camera. Since the geometry and the distribution of light deployed onto the camera can be easily reconstructed analytically for a muon of given energy and direction, muon rings are a powerful tool for monitoring the behaviour of crucial properties of an imaging telescope such as the point-spread-function and the overall light collection efficiency.

In this contribution we present the possibility of using the analysis of muon ring images as calibrator for the ASTRI SST-2M prototype point spread function.}


\begin{document}
\maketitle

%Begin a section.
\section{Introduction}

ASTRI ("Astrofisica con Specchi a Tecnologia Replicante Italiana") is a Flagship Project financed by the Italian Ministry of Education, University and Research, and led by the Italian National Institute of Astrophysics, INAF.
The Project \cite{bib:astri} is strictly linked to the Cherenkov Telescope Array, CTA \cite{bib:CTA}, since it is currently developing an end-to-end prototype \cite{bib:pareschi} of the Small Size class Telescopes with wide field of view (full FoV $\sim$9.6$^{\circ}$) aimed to observe the highest energy range investigated by CTA.

The telescope, named ASTRI SST-2M, will be placed at the INAF "M.G. Fracastoro" observing station in Serra La Nave, 1735 m a.s.l. on the Etna Mountain near Catania, Italy \cite{bib:maccarone}. The installation is foreseen in middle 2014 and the data acquisition is scheduled to start soon after.

The ASTRI SST-2M telescope is characterized by two innovative technological solutions, for the first time adopted together in the design of Cherenkov telescopes: the optical system is arranged in a dual-mirror configuration \cite{bib:canestrari} and the camera at the focal plane is composed by a matrix of multi-pixel Silicon Photo Multipliers \cite{bib:catalano}. A careful calibration phase is then required to probe the technological solutions adopted and to verify the ASTRI SST-2M telescope expected performance \cite{bib:bigongiari}.

The verification tests will be conducted soon after the installation of ASTRI SST-2M in Serra La Nave making use of different approaches and thanks also to auxiliary systems, inner \cite{bib:catalano} and outer of the telescope \cite{bib:maccarone}. Further similar tests, together with a technical and scientific calibration plan, will be performed during the regular data-taking; among them, the telescope performance could be monitored through the study of muon ring images.

In this paper we study the possibility of using the analysis of muon ring images as calibrator for the ASTRI SST-2M prototype point spread function (PSF).

\section{Muons in Calibrations}

Muons reach the ground in a steady flux. For energy higher than $\sim$ 5 GeV they produce Cherenkov radiation that can be observed by the Imaging Atmospheric Cherenkov Telescopes (IACTs). They are observable also with heavily \-co\-ve\-red sky, as the Cherenkov photons collected are produced in the last part of the path (few hundreds of meters). Muons are single superluminal particles and they radiate in a cone producing a characteristic ring image in the telescope \-ca\-me\-ra, with a typical radius of about 1$^{\circ}$, when incoming with a small impact parameter.\\ 
In case of single telescope, Cherenkov light from cosmic-ray muons is a significant source of background and deteriorates the sensitivity of the IACTs, while for arrays of telescopes the muon background is almost totally eliminated at the trigger level. However, the analysis of ring images generated from high-energy muons is a valuable diagnostic tool to monitor the behaviour of crucial properties of telescopes such as the PSF and the overall light efficiency, as widely studied and described by several authors and applied onto several Cherenkov telescopes as Whipple, VERITAS, MAGIC and H.E.S.S.. \cite{bib:Rose,bib:Humensky,bib:Meyer,bib:Bolz}. In fact, from theo\-re\-ti\-cal formulas \cite{bib:Rovero,bib:vacanti} and from simulations it is possible to reconstruct the number of photons at the telescope pupil and the intrinsic broadening of the Cherenkov emission. The number of detected photoelectrons and the PSF of the image can be used to check the system efficiency and the muon ring broadening due to mirror aberration and defocusing.
 
\begin{figure}[h]
  \centering
  \includegraphics[width=0.475\textwidth]{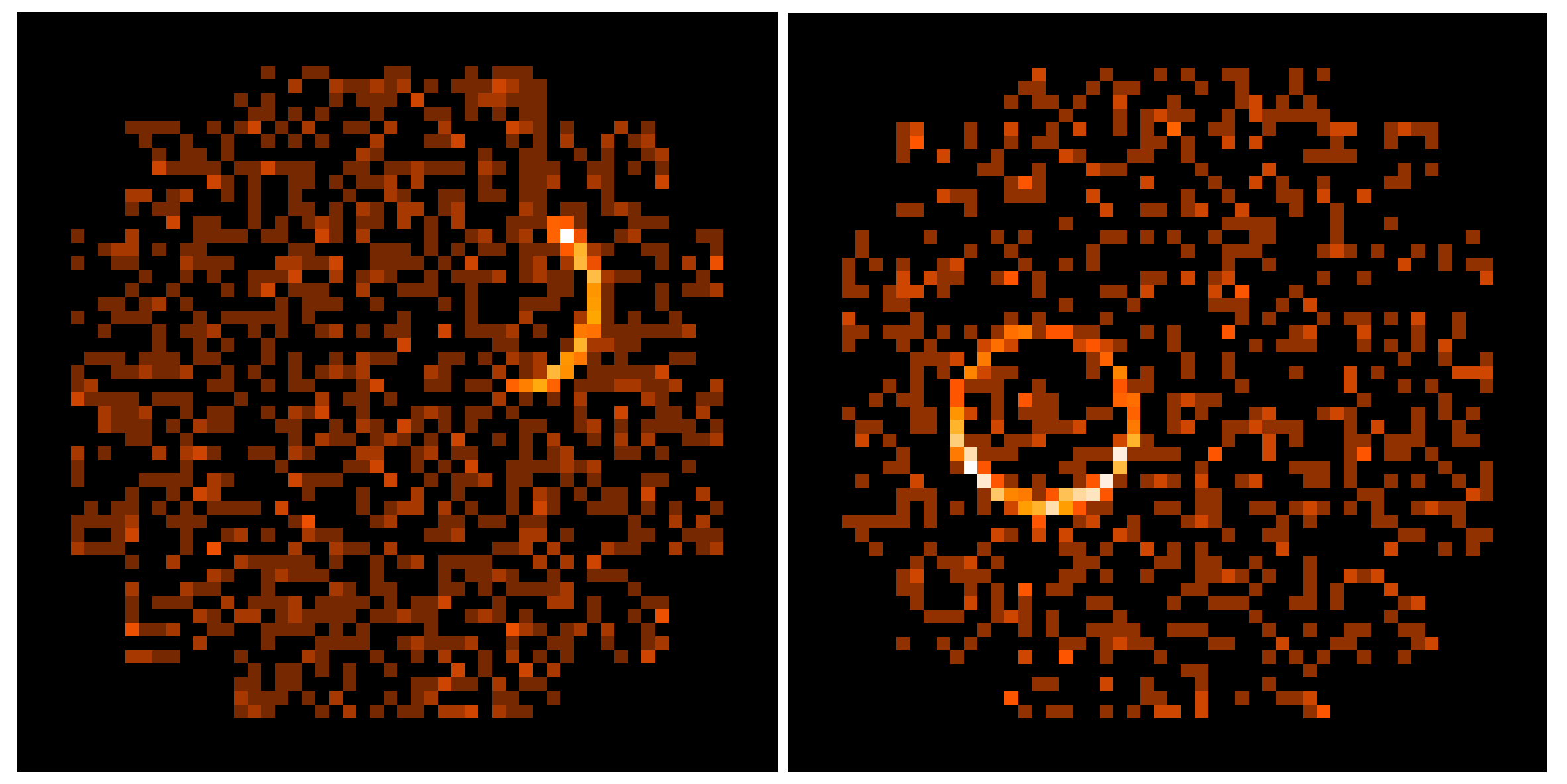}
  \caption{How the ASTRI SST-2M camera would visualize Cherenkov light from muons embedded in high level of night sky background (more than a quarter Moon). Left and right panels: muon events simulated with Energy 73 GeV and 21 GeV, arrival zenith angle of 1.9$^{\circ}$ and 1.7$^{\circ}$, impact point at $\sim$2 m and $\sim$1.6 m from the center of the telescope, respectively.}
  \label{muoni}
 \end{figure}

Figure \ref{muoni} shows how the Cherenkov light produced by two separate muon events would be seen by the ASTRI SST-2M camera. The events have been simulated with a similar arrival direction (the angle with respect to the telescope optical axis and pointing to the zenith direction is $\sim$1.8$^{\circ}$) and with primary energy of 73 GeV and 21 GeV. The left panel presents the image of muon whose impact point is at the edge of the ASTRI SST-2M primary mirror, while the right panel is relative to an event with impact point 1.6 m from the center of the telescope.

Although no real data are available at the moment for the ASTRI SST-2M telescope, it is possible to outline the steps that would be performed once real data will be acquired. This will include:
\begin{itemize}
  \item selection of the muon ring image;
  \item estimation of the center and radius of the muon ring;
  \item calculation of the width of the muon ring;
  \item reconstruction of the light intensity distribution along the muon ring.
\end{itemize}
The final point of this procedure will concern the comparison with a suitable set of simulated data opportunely crea\-ted in a specific database to take into account both technica\-lities of our ASTRI SST-2M telescope and characteri\-stics of the site where it will be installed. All the above mentioned steps of the procedure will be tested on this \-si\-mu\-la\-tion database.

\section{Simulation and Method of Analysis}

The main technical features of the ASTRI SST-2M telescope \cite{bib:pareschi} to be considered for our muon events simulation can be briefly summarized as follows: the collecting area of the primary mirror whose radius is 2.153 m; the logical pixel of the camera at the focal plane is approximately of 0.17$^{\circ}$; they are arranged in 37 Photon Detection Modules (PDM) of 8x8 logical pixels each; the full field of view corresponds to 9.6$^{\circ}$. With regard to the Serra La Nave ob\-ser\-ving station where the prototype will be installed \cite{bib:maccarone}, the site altitude is 1735 m a.s.l. and the magnetic field is equal to 26.84 ${\mu}$T and 36.04 ${\mu}$T in its horizontal and vertical components, respectively.

Taking into account such features and using CORSIKA.6970 \cite{bib:Heck} with IACT/ATMEXT.1.44 version, we simulated a first set of muon events with arrival directions uniformly sorted within a cone of 2.5$^{\circ}$ opening angle around the telescope axis in order to have events that can produce complete rings at the focal plane. The muon starting altitude was defined at 717.25 g/cm$^{2}$ correspondent to 3 km a.s.l., about 1.2 km above the Serra La Nave altitude. The U.S. Standard atmosphere model was considered.

The CORSIKA data-files were therefore passed through the telescope simulation code which ray-traces photons from the primary mirror to the second one till the pupil of the focal plane camera where they are eventually converted in a photoelectron list forming the image to be analysed.

The flux of muons expected to produce complete ring are computed from \cite{bib:sanuki} assuming a maximum zenith angle of 2.5$^{\circ}$ and a maximum distance of the impact point equal to the primary mirror radius.  We find a total ($\mu^{-}$ and $\mu^{+}$) rate of 1.5 muons/sec with a contribution of about 50\% from each component. The simulation data-set here analysed is relative only to $\mu^{-}$ muons with energies from 6 GeV to 80 GeV, subdivided in three different bands. The number of events in each energy range was computed using the spectral parameter obtained fitting the $\mu^{-}$ spectrum in \cite{bib:sanuki} with power laws whose differential spectral indices are shown in Table~\ref{table_dataset}) where the number of CORSIKA simulated events is also presented.

\begin{table}[h]
\begin{center}
\begin{tabular}{|c|c|c|c|}
\hline
Energy & Spectral & Simulated & Analysed\\
Range [GeV] & Index & Events & Events\\
\hline
6-10& -2.16& 100000& 14938\\
\hline
10-20& -2.46& 63000& 17881\\
\hline
20-80& -2.71& 25000& 11566\\
\hline
\end{tabular}
\caption{The muon simulated data-set. Last column refers to the number of events survived to the selection (see text) and used for the final evaluation of the procedures}
\label{table_dataset}
\end{center}
\end{table}

Our main interest is to probe the feasibility of using muon ring images as calibrators for the ASTRI SST-2M telescope and consequently to test the specific analysis procedure just after the application of the cleaning algorithm that will be implemented in the ASTRI data analysis system \cite{bib:antonelli}. Therefore, to address the work directly to our target, in this phase no night sky background was added to the muon events.

Events with a number of photoelectrons $>$40 and with a number of fired pixels $>$20 are selected for the analysis. For each single muon event we evaluated:
\begin{itemize}
  \item the center of the muon ring fitting with a circle the focal plane image;
  \item the radius (\emph{Radius}) and the width (\emph{ArcWidth}) of the ring fitting with a Gaussian the distribution of the number of detected photoelectrons versus the distance: the mean of the Gaussian gives the \emph{Radius} and the standard deviation $\sigma$ gives the \emph{ArcWidth};
  \item the light intensity distribution along the muon ring using the formula in \cite{bib:vacanti} relative to the azimuthal photon distribution for events with impact parameter smaller than the primary mirror radius.
\end{itemize}

After the analysis we applied a further selection con\-si\-de\-ring events with \emph{Radius}$<$2$^{\circ}$ and \emph{ArcWidth}$<$0.2$^{\circ}$ in agreement with previous work \cite{bib:Meyer}.

\section{Analysis Results}
We checked the precision of the reconstructed muon direction and impact point comparing the results of our analysis with the known input parameters of CORSIKA simulations. We find that the muon direction is reconstructed with an error of 5\% and the impact point with an indetermination of about 50\%.

\begin{figure}[!t]
 \centering
 \includegraphics[width=0.5\textwidth]{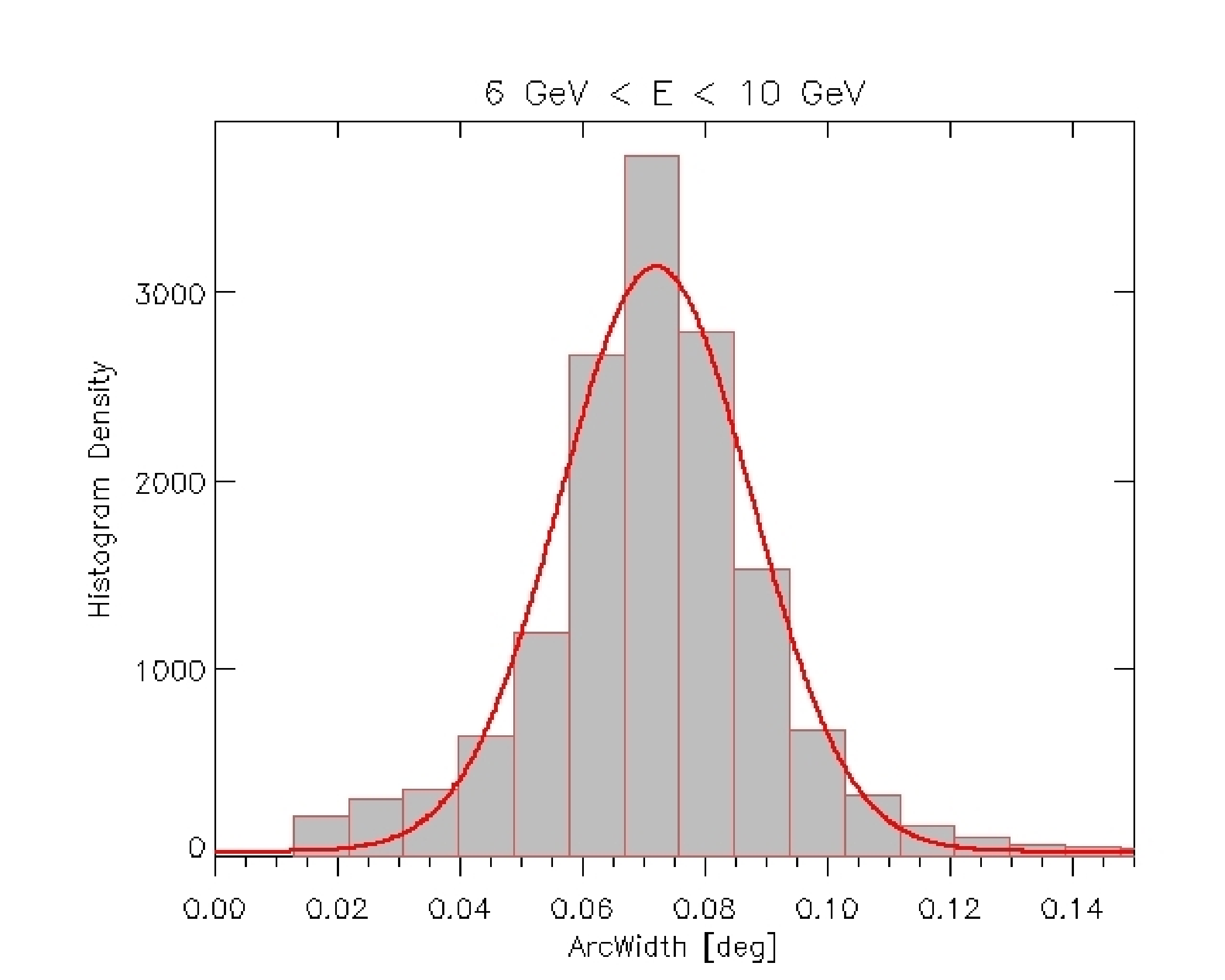}
 \includegraphics[width=0.5\textwidth]{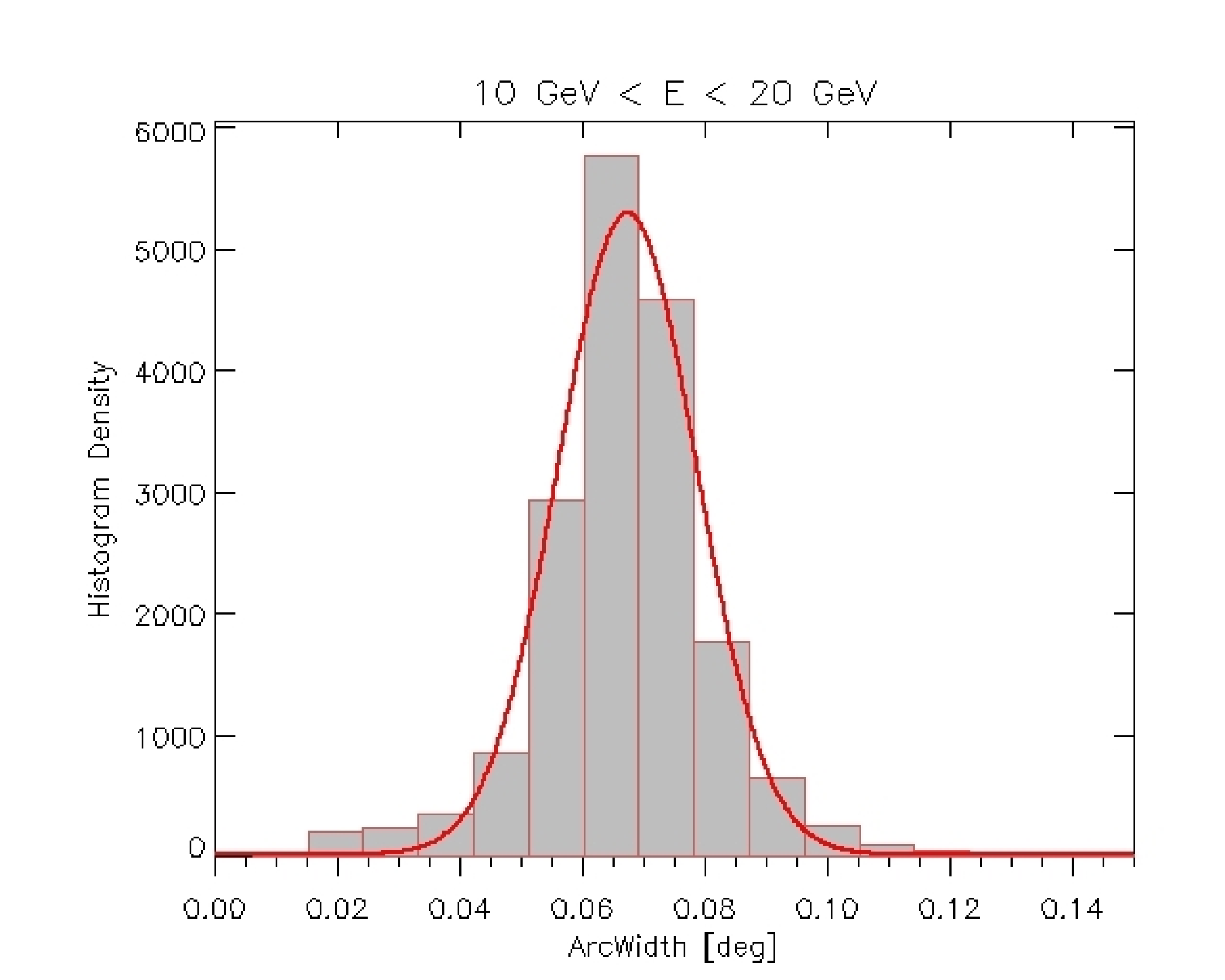}
 \includegraphics[width=0.5\textwidth]{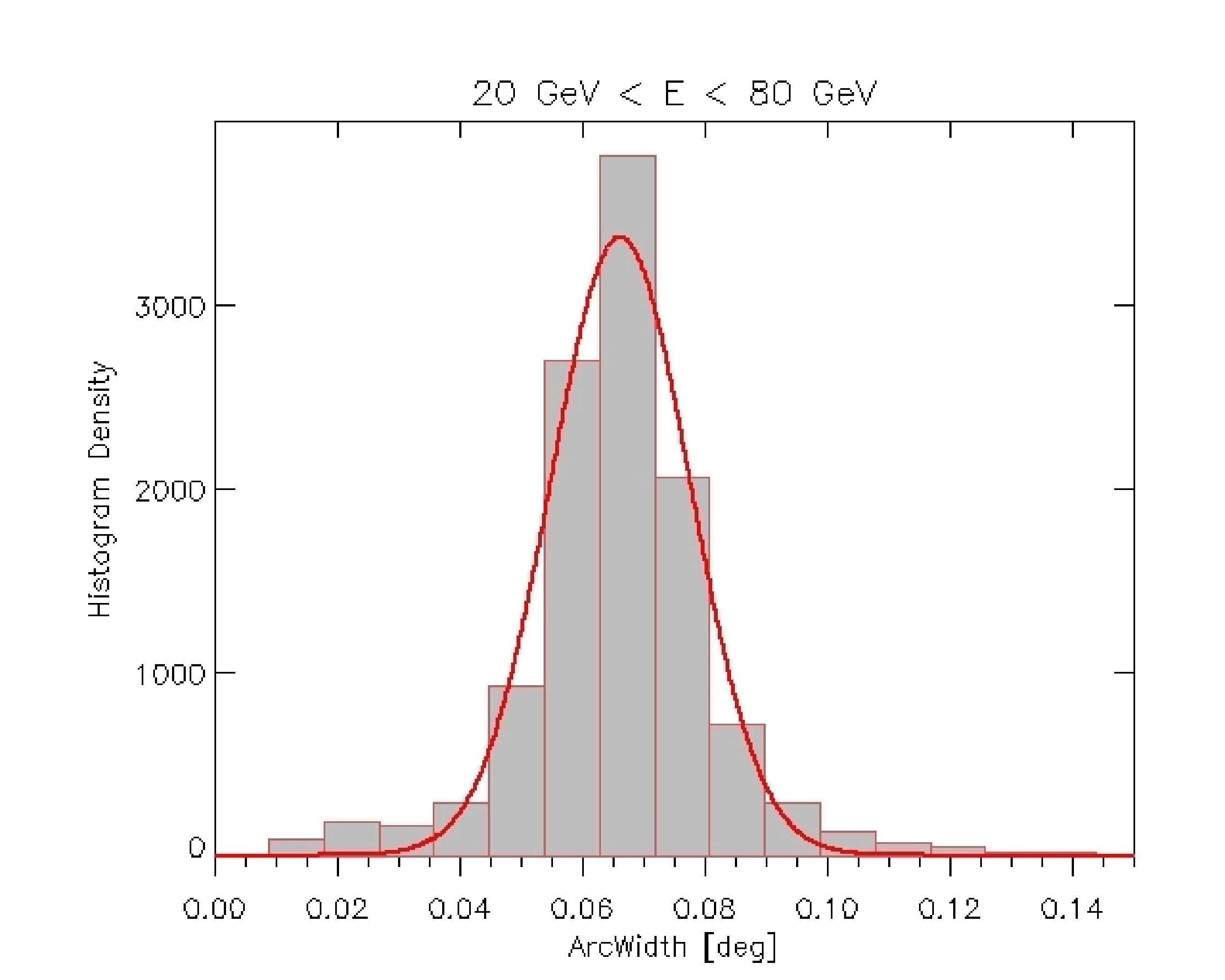}
 \caption{\emph{ArcWidth} distributions for the three energy ranges adopted in the analysis}
 \label{fig:2}
 \end{figure}

We plot in Fig.~\ref{fig:2} the distributions of the reconstructed \emph{ArcWidth} for the three energy ranges adopted in the analysis. The curves were fitted with Gaussian functions whose central values are reported in Table 2 with the errors given by the Gaussian standard deviation. The \emph{ArcWidth} distribution, that is used to estimate the telescope PSF, is peaked at 0.07$^{\circ}$. This value is dominated by the optics PSF that is 0.06$^{\circ}$ as obtained by the simulator, and corresponds to the radius at which the 80\% of the energy is encircled; this spread introduced by the broadening due to the involved physical processes is negligible \cite{bib:Meyer}.

\begin{figure}[!t]
 \centering
 \includegraphics[width=0.5\textwidth]{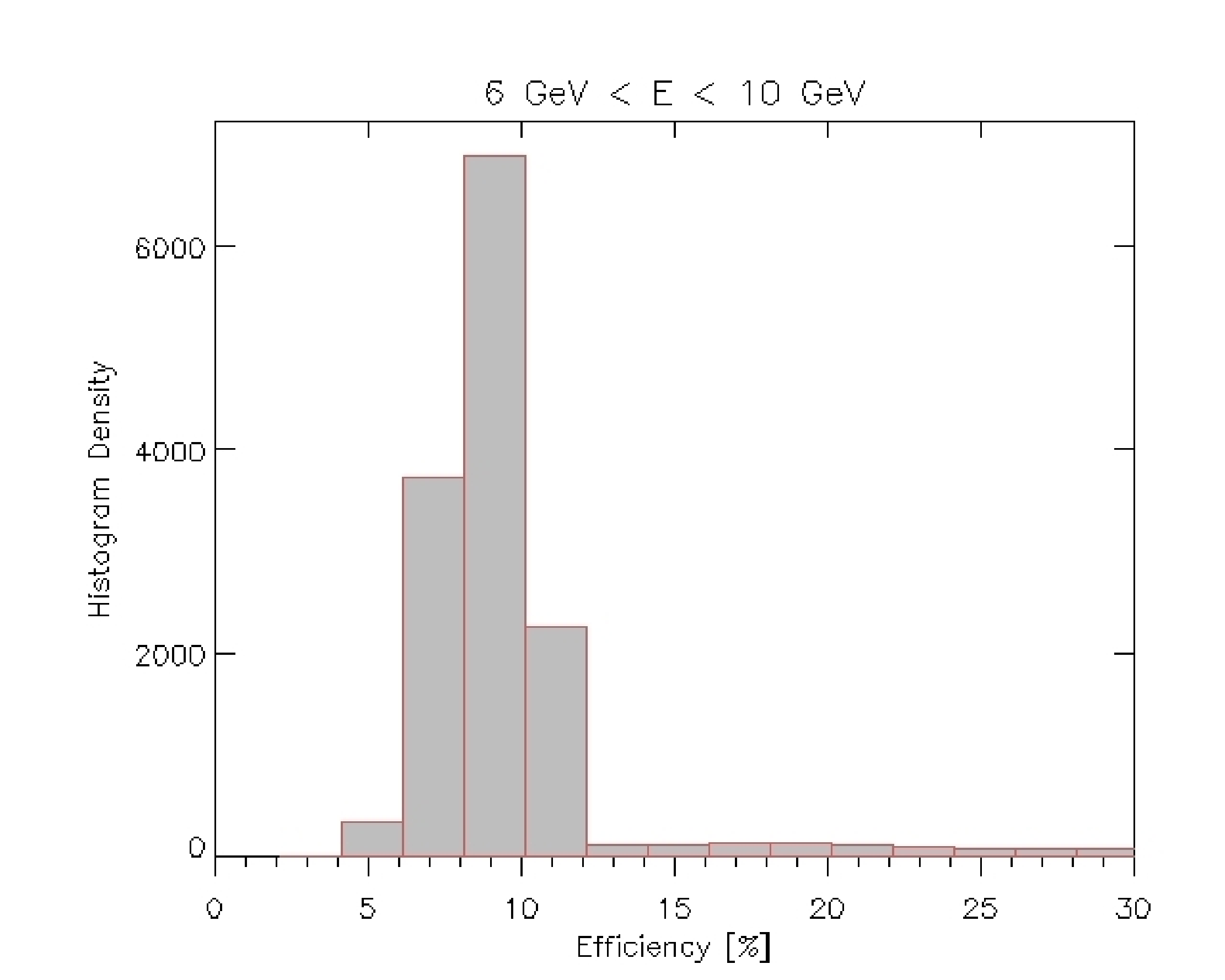}
 \includegraphics[width=0.5\textwidth]{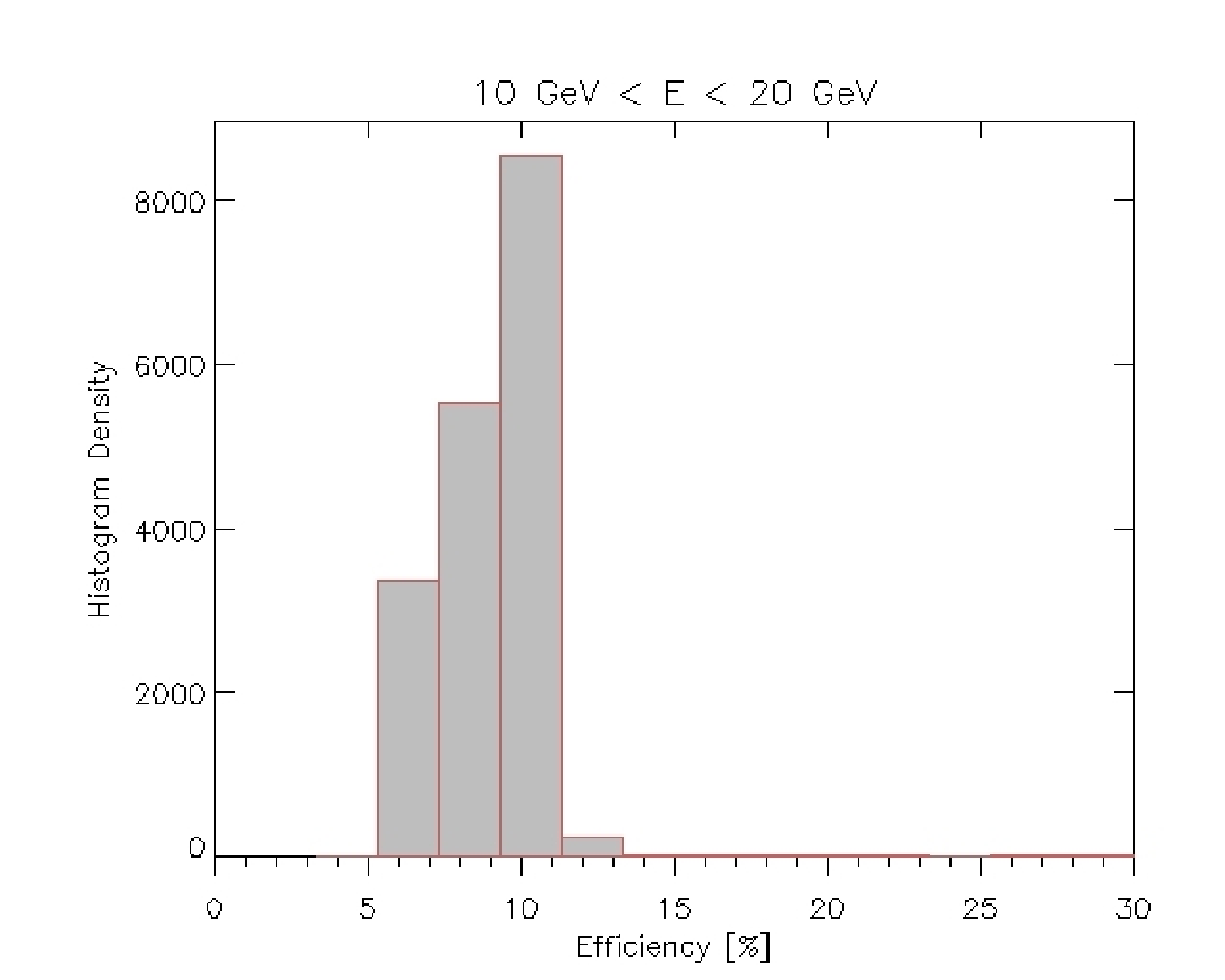}
 \includegraphics[width=0.5\textwidth]{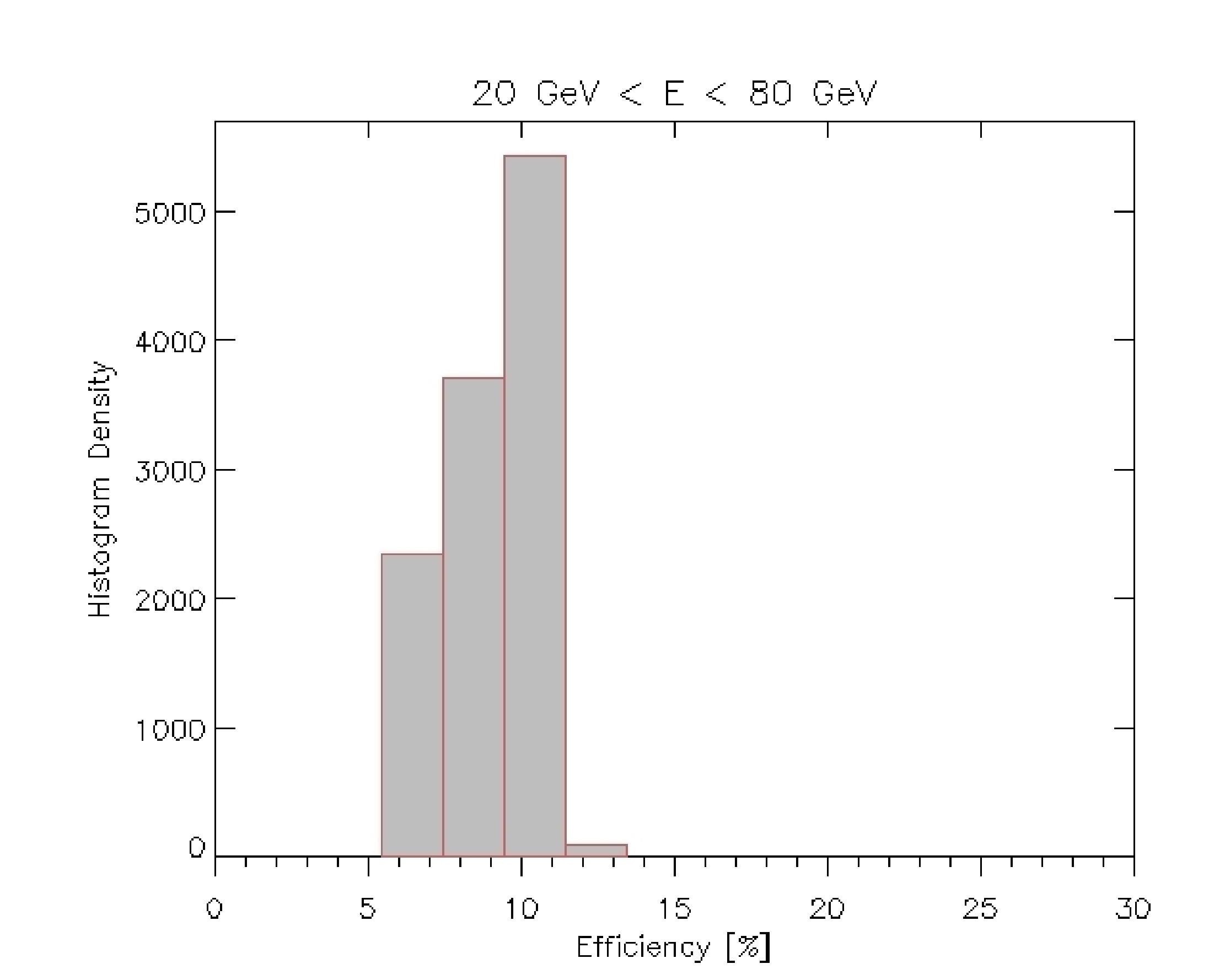}
 \caption{Overall detection Efficiency distributions for the three energy ranges adopted in the analysis}
 \label{fig:3}
 \end{figure}

Moreover we evaluated the overall efficiency of the telescope, plotted in Fig.~\ref{fig:3}, computing the ratio between the detected photoelectrons and the photons at the telescope pupil evaluated from the simulations. We found that the telescope efficiency is about 10\%.

\begin{table}[t]
\begin{center}
\begin{tabular}{|c|c|c|}
\hline
Energy &\emph{ArcWidth} &Telescope Eff\\
$[GeV]$ &$[Deg]$ &$[\%]$\\
\hline
6-10& 0.07$\pm$0.01& 9.4$\pm$1.6\\
\hline
10-20& 0.07$\pm$0.01& 8.8$\pm$1.5\\
\hline
20-80& 0.07$\pm$0.01& 9.4$\pm$1.7\\
\hline
\end{tabular}
\caption{\emph{ArcWidth} and Efficiency mean values for each energy range computed fitting a Gaussian to each curve in Fig.~\ref{fig:2} and Fig.~\ref{fig:3}. Errors are given by the standard deviation of the Gaussian.}
\label{table_results}
\end{center}
\end{table}

\section{Conclusions}
We implemented a method to analyse the muon ring images and to estimate the center and radius of the ring, as well as reconstruct the light intensity distribution along the muon ring.
We simulated a set of muon events reaching the primary mirror of ASTRI SST-2M telescope in order to check the feasibility of using muons as calibrator.  
We found no difference in the ArcWidth distribution for different energy ranges.
We evaluated the overall PSF from the width of the analysed ring and we obtained values in agreement with our knowledge of the optics PSF. 
We found that we are able to measure the PSF with a precision of about 14\%. Moreover we evaluated the precision of reconstructing the muon direction and the impact point: we found that the muon direction is reconstructed with an error of 5\% and the impact point with an indetermination of about 50\%.

We calculated the value of the overall telescope efficiency as the ratio between the detected photoelectrons and the photons at the pupil to test the validity of our method and we found that it is about 10\%, in agreement with the simulation results of gamma induced showers. 

Next step will be the implementation of a method to compute the efficiency using the correlation between the number of detected photoelectrons and the value of \emph{Radius}; a comparison with simulated data will be performed.

\vspace*{0.5cm}
\footnotesize{\bf Acknowledgment:}{This work was partially supported by the ASTRI Flagship Project financed by the Italian Ministry of Education, University, and Research (MIUR) and led by the Italian National Institute of Astrophysics (INAF). We also acknowledge partial support by the MIUR Bando PRIN 2009.}

\FloatBarrier


\begin{thebibliography}{}

\bibitem{bib:astri} G. Pareschi et al., [The ASTRI Collaboration], (in preparation)

\bibitem{bib:CTA} B.S. Acharya et al. [The CTA Consortium], {Astroparticle Physics} 43, 3 (2013)

\bibitem{bib:pareschi} G. Pareschi et al., 33rd ICRC, 2013, id 466, these proceedings

\bibitem{bib:maccarone} M.C. Maccarone et al., id$\textunderscore$110, these proceedings

\bibitem{bib:canestrari} R. Canestrari et al., id$\textunderscore$468, these proceedings

\bibitem{bib:catalano} O. Catalano et al., id$\textunderscore$111, these proceedings

\bibitem{bib:bigongiari} C. Bigongiari et al., id$\textunderscore$564, these proceedings

\bibitem{bib:Rose} H.J.Rose, 24rd ICRC, 1995, 3, 464  

\bibitem{bib:Humensky} T.B.Humensky, astro-ph/0507449 

\bibitem{bib:Meyer} M.Meyer et al., AIP Conf.Proc. 745 (2005) 774-778

\bibitem{bib:Bolz} O.Bolz, Verhandlungen der Deutschen Physikalischen Gesellschaft, 38, 39 (2003)

\bibitem{bib:Rovero} A.C.Rovero et al., Astroparticle Phys. 5 (1996) 27

\bibitem{bib:vacanti} G. Vacanti et al., {Astroparticle Physics}, 1994, 2, {1-11}

\bibitem{bib:Heck} D. Heck et al., Forschungszentrum Karlsruhe Report No. FZKA 6019 (1998)

\bibitem{bib:sanuki} T. Sanuki et al., Phys. Letters B, 541, Issue 3-4, 234 (2002)

\bibitem{bib:antonelli} L.A. Antonelli et al., id$\textunderscore$925, these proceedings

\end{thebibliography}
\end{document}